\documentclass[12pt]{article}
\usepackage{amsmath}
\usepackage{graphicx}

\begin{document}
\begin{center}
{\large\bf  Formulae for estimating average particle energy loss
due to Beamstrahlung in supercolliders}
\end{center}
\begin{center}
{S.A. Nikitin\footnote{Sergei Nikitin\\nikitins@inp.nsk.su}\\ BINP SB RAS, Novosibirsk, RF}
\end{center}
\section*{Abstract}
Based on simplified models, formulae for determining particle energy losses due to Beamstarhlung in supercolliders are obtained. The developed semi-analytical approach can be useful for estimating the parameters of colliding beams under various conditions without using special beam-beam simulation codes.

\section{Introduction}
One of the serious problems of modern projects of lepton high energy circular supercolliders \cite{FCC,CEPC} is the radiation of accelerated particles in the collective field of an oncoming bunch (Beamstruhlung). First of all, this effect can lead to a significant increase of bunch length and energy spread. As a result, the collider luminosity and the energy resolution of experiments reduce. The study of Beamstruhlung (BS) and optimization of beam parameters is carried out using the special Beam-Beam simulation codes. As a rule, this work requires increased computer resources and is carried out by the code creator. In this paper, we set the goal to derive formulae for the numericalß estimation of energy losses for BS  in order to simplify the initial optimization of the accelerator parameters, making this stage more accessible, i.e. without performing  beam-beam simulation.

\section{Collinear-collision-based model}
For our aims, we use the formula by K.Takayama for the potential of a Gaussian ellipsoidal bunch  of  $N$  particles at rest \cite{Tak}. In Lab frame $(x,y,z)$, that bunch moves along the $z$  axis. In the accompanying system $(x',y',z')$, this potential can be defined as \cite{LN}
	\begin{equation}\label{TP}
\Phi'(x',y',z')=\frac{eN}{\sqrt{\pi}}\int_{0}^{\infty}\frac{\exp{\left\{-\frac{x'^2}{2\sigma_x^2+t}-\frac{y'^2}{2\sigma_y^2+t}-\frac{z'^2}{2\gamma^2\sigma_z^2+t}\right\}}}{\sqrt{(2\sigma_x^2+t)(2\sigma_y^2+t)(2\gamma^2\sigma_z^2+t)}}dt,		
\end{equation}
where  $\sigma_x,\sigma_y,\sigma_z$ are the transverse ($x,y$) and longitudinal ($z$) beam sizes in Lab;   $\gamma$ is the Lorentz factor which is the same for both the colliding beams at the beam energy $E_0$ in Lab. Here, it is assumed that the axes of the same name of the $(x',y',z')$ and $(x,y,z)$ systems are parallel to each other, and at some point in time these systems coincide with the origin. We will neglect the relative motion of charges inside the  bunches.

\begin{figure}[htb]
\centering
\includegraphics*[width=120mm]{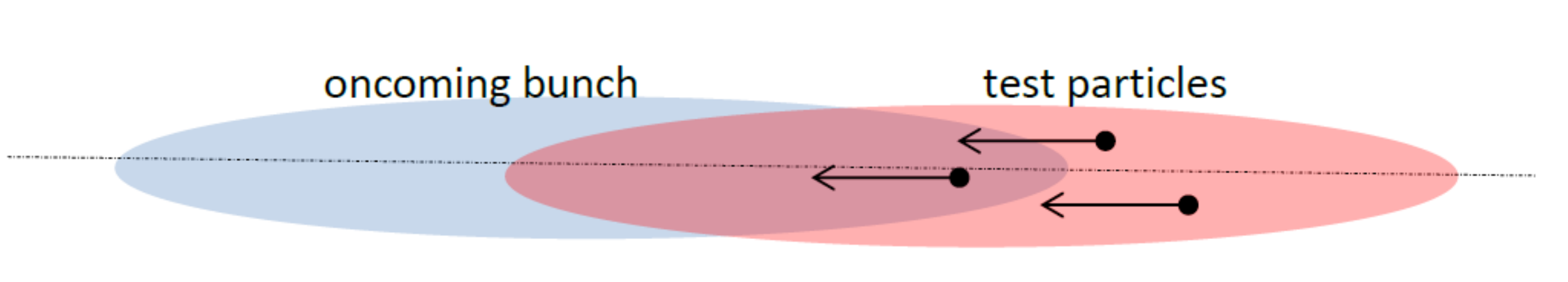}
%\vspace{-15mm}
\caption{ In the collinear model, the test particles move parallel to the common axis for colliding 
beams (head-on model).}
\label{bsf1}
\end{figure}

In the general case, the amount of energy radiated by an electron while moving with velocity  $\vec\beta c$ in the external electrical ($\vec E$ ) and magnetic ($\vec H$ ) fields can be found from the equation \cite{LL}:

\begin{equation}\label{eLL}
W=\frac{2}{3}\frac{e^4}{m^2c^3}\int_{-\infty}^{\infty}\frac{\{\vec E+[\vec \beta\vec H]\}^2-(\vec E\vec\beta)^2}{1-\vec \beta^2}d\tau		
\end{equation}
where the integral is taken over time. Let a test particle move in the rest system of the oncoming bunch along the $z'$ axis with arbitrary transverse coordinates $x = x'$ and $y = y'$  (Fig.\ref{bsf1}). Due to the very short interaction length and small energy losses, we will neglect the curvature of the trajectory, assuming that these coordinates as well as particle energy do not notably change. The energy loss due to radiation is determined through the integral over a time $\tau$   in the oncoming bunch rest frame:
\begin{equation}\label{eLLr}
W_{rest}=\frac{2}{3}\frac{e^4}{m^2c^3}\int_{-\infty}^{\infty}\gamma^{\prime 2} E_{\perp}^{\prime 2}d\tau,		
\end{equation}
where $E_{\perp}^{\prime 2}=E_{x'}^{\prime 2}+E_{y'}^{\prime 2} $ with
\begin{equation}\label{Exy}
\begin{split}
			E'_{x'} =-\frac{\partial \Phi'}{\partial x'}, \\
			E'_{y'} =-\frac{\partial \Phi'}{\partial y'} 
\end{split}
\end{equation}
being the electrical field components calculated from (\ref{TP}).  The relativistic factor of test particles in the rest frame of oncoming bunch is $\gamma'\approx2\gamma^2$. In Lab, the value of energy loss is 
\begin{equation}\label{eLab}
W_{Lab}\approx\frac{W_{rest}}{2\gamma}.		
\end{equation}

Integrating in (\ref{eLLr}) over $z'\approx\tau c$,  and then averaging the result over $x'$ and $y'$ with the appropriate Gaussian  distribution functions, and finally using (\ref{eLab}), we obtain the relative BS loss for the case of  pure head-on collision:
\begin{equation}\label{Ubs0}
U_{bs}^{(head-on)}=\frac{W_{Lab}}{E_0}=\frac{16 r_e^3 N^2\gamma^2}{3\sqrt{\pi}}\int_0^\infty\int_0^\infty\frac{dtdt'}{\sqrt{4\gamma^2\sigma_z^2+t+t'}}\left\{\frac{\sigma_x^2}{\sqrt{g_x^3g_y}}+\frac{\sigma_y^2}{\sqrt{g_y^3g_x}}\right\},		
\end{equation}
\begin{equation*}\label{gx}
g_x(t,t')=(2\sigma_x^2+t)(2\sigma_x^2+t')+2\sigma_x^2(4\sigma_x^2+t+t'),
\end{equation*}
\begin{equation*}\label{gy}
g_y(t,t')=(2\sigma_y^2+t)(2\sigma_y^2+t')+2\sigma_y^2(4\sigma_y^2+t+t').
\end{equation*}

\begin{figure}[htb]
\centering
\includegraphics*[width=110mm]{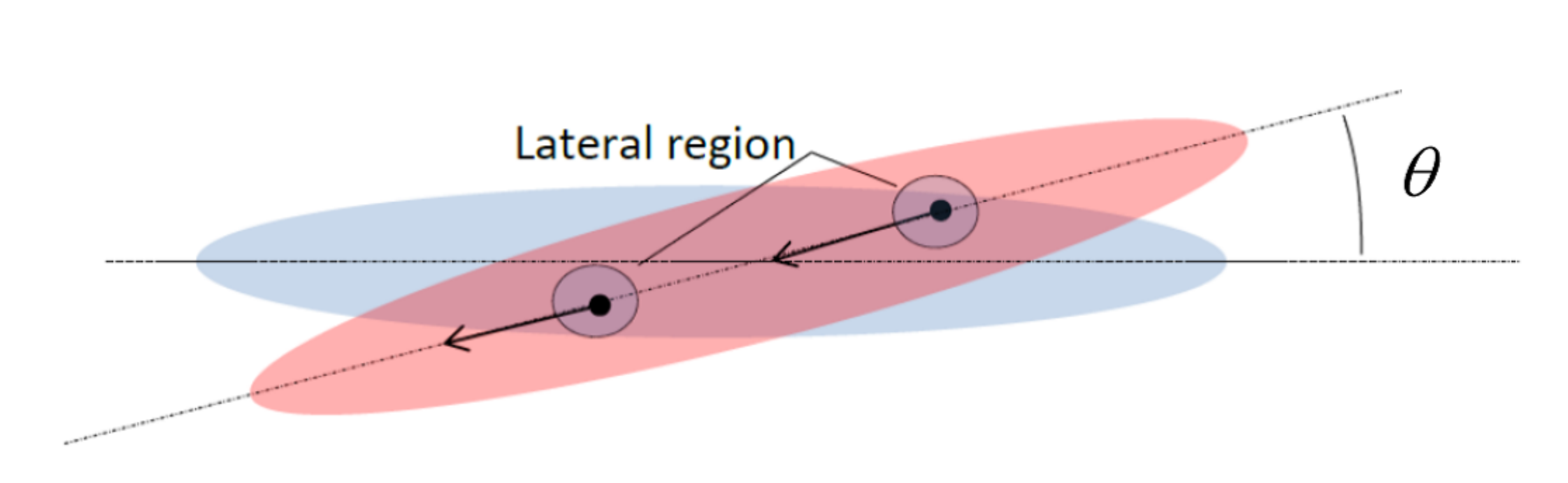}
%\vspace{-15mm}
\caption{ Non-zero crossing angle reduces the interaction length. There is a need to take into account the radiation of axial test particles in the lateral regions of  the oncoming beam.}
\label{bsf2}
\end{figure}

High luminosity design of the FCCee and CEPC projects  is based on the Crab Waist scheme \cite{Rai,MZ} of beam-beam interaction which implies  a non-zero crossing angle   $\theta\ll1$  Fig.(\ref{bsf2}). As a result, the characteristic size of the interaction region differs from the longitudinal size of the beam  and approximately equal to 
\begin{equation}\label{sig}
\frac{\sigma_z}{\sqrt{1+\Psi^2}}\approx\frac{2\sigma_x}{\theta},		
\end{equation}
where  $\Psi=\frac{\sigma_z}{\sigma_x}\tan{\frac{\theta}{2}}\gg 1$  is a so-called Piwinski angle ($\sigma_z\gg\sigma_x$) .
Given the relation (\ref{sig}), the formula (\ref{Ubs0}) for estimating the relative energy loss in Lab is modified and takes a form \cite{LN}: 
\begin{equation}\label{Ubs1}
U_{bs}^{(1)}=\frac{W_{Lab}}{E_0}\approx \frac{16 r_e^3 N^2\gamma^2}{3\sqrt{\pi}\sqrt{1+\Psi^2}}\int_0^\infty\int_0^\infty\frac{dtdt'}{\sqrt{4\gamma^2\sigma_z^2+t+t'}}\left\{\frac{\sigma_x^2}{\sqrt{g_x^3g_y}}+\frac{\sigma_y^2}{\sqrt{g_y^3g_x}}\right\}.
\end{equation}

In practical numerical calculations,  the term $4\gamma^2\sigma_z^2+t+t'$  can be substituted by $4\gamma^2\sigma_z^2$   without  any noticeable loss of accuracy due to the very large superiority of the bunch length over the transverse dimensions.		

We call the considered approach Model 1. This approach  is true with an accuracy, determined by amount of the additional losses depending on the crossing angle. In particular, for particles placed on the axis of their own bunch, the field of the counter bunch in the model is zero. In fact, these particles pass through the lateral regions of the oncoming bunch with non-zero transverse fields and therefore radiate. A further approximation in order to take into account these additional losses is the model described below.

\section{Non-collinear collision: Model 2}
\begin{figure}[htb]
\centering
\includegraphics*[width=100mm]{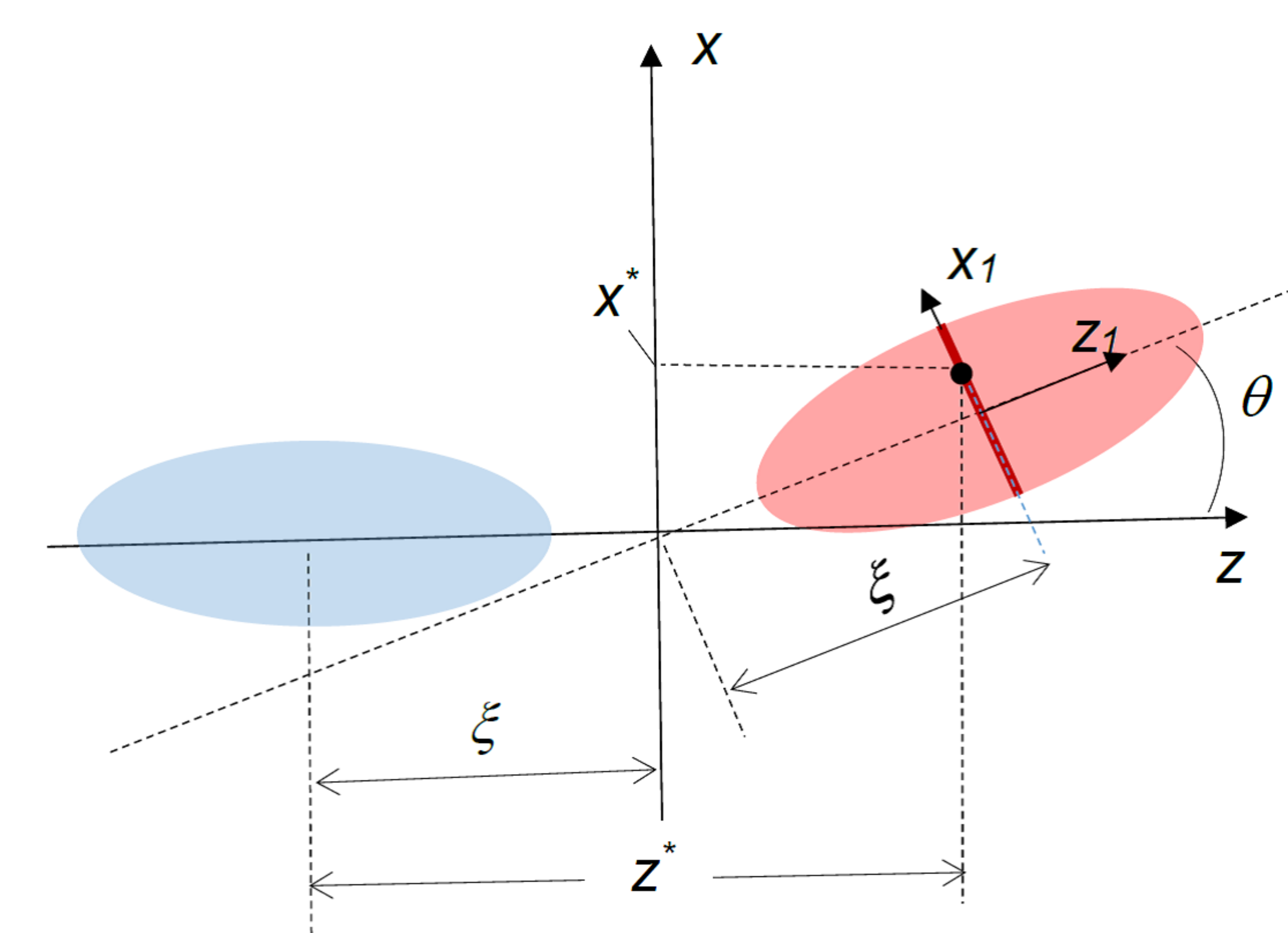}
%\vspace{-15mm}
\caption{Parameters for describing the disposition of a test particle, belonging to the central cross section of the beam ($z_1=0$), relative to the oncoming bunch in Lab.}
\label{bsf3}
\end{figure}

Consider a more detailed model at non-zero crossing angle (Fig.\ref{bsf3}). This model allows taking into account radiation of the axial test particles in the 'lateral regions' of the oncoming bunch. Beside, in this approach, one can naturally involve a contraction of interaction length. 

First, we define the parameters describing location of the test particles relative to the coordinate system associated with the oncoming beam, through their coordinates in Lab  as shown in Fig.3. Let $\xi$  be  $z$ -  coordinate of the center of oncoming bunch, and $x_1,z_1$ be the coordinates in the axes related to the test particle bunch (all these quantities are treated in Lab). Then the parameters of the particle location of interest to us are as follows: 
\begin{equation}\label{coor1}
z^*\approx 2\xi-x_1\theta+z_1,\quad
x^*\approx \xi\theta+x_1+z_1\theta.
\end{equation}
In the system of the resting oncoming bunch,  these coordinates will be
\begin{equation}\label{coor2}
z^{\prime*}\approx \gamma(2\xi-x_1\theta+z_1),\quad 
x^{\prime*}\approx \xi\theta+x_1+z_1\theta.
\end{equation}
To simplify the task, we put $z_1=0$. It means that we will search for the energy loss averaged over the central cross section of the bunch with test particles. We express the electric fields  (\ref{Exy}) in the equation    (\ref{eLLr})  in terms of the coordinates of the test particles, taking into account (\ref{coor2}). Then we obtain  the average BS loss of these particles in the oncoming beam in the framework of Model 2:
\begin{equation}\label{Ubs2}
U_{bs}^{(2)}\approx \frac{4 r_e^3 N^2\gamma}{3\sqrt{\pi}\sigma_z}\int_0^\infty\int_0^\infty\frac{dtdt'}{\Omega\sqrt{(2\sigma_x^2+t)(2\sigma_x^2+t')}}\left[\frac{\sigma_x^2}{g_x\sqrt{g_y}}\left(1+\frac{\theta^2\sigma_z^2}{8\sigma_x^2\Omega^2}\right)+\frac{\sigma_y^2}{\sqrt{g_y^3}}\right].
\end{equation}
The functional
\begin{equation}\label{omeg}
\Omega(t,t')=\sqrt{1+\left(\frac{\theta}{2}\right)^2\left(\frac{1}{2\sigma_x^2+t}+\frac{1}{2\sigma_x^2+t'}\right)\sigma_z^2}
\end{equation}
represents a generalized Piwinski factor. When $t=t'=0$, it takes the known form: 
\begin{equation*}\label{omeg1}
\Omega(0,0)=\sqrt{1+\left(\frac{\sigma_z\theta}{2\sigma_x}\right)^2}\approx\sqrt{1+\Psi^2}.
\end{equation*}

The average value of the particle energy loss $U_{bs}^{(2)}$  in the central section, obtained taking into account the kinematic features at a non-zero crossing angle,  is, apparently,  an upper estimate. In any other cross sections, the losses will be lower, since they occur in less dense regions and, therefore, in smaller fields. $U_{bs}^{1)}$ is the BS loss found in the most evident  model of head-on collision and then  modified with the conventional Piwinski factor. This value do not include the losses of particles moving on the bunch axis. Therefore, it should be borne in mind that Model 1 can underestimate energy losses to a greater extent than Model  2 overstates them.

The numerical evaluation of double integrals in the formulae (\ref{Ubs1}), (\ref{Ubs2}) is available using, for instance, the Maple 14 computing platform.  Calculation of the energy loss for BS takes about 1 minute. 

\section{Length of equivalent magnet}
The formulae obtained can be useful for estimating the influence of BS on the formation of the longitudinal beam size and energy spread in supercolliders \cite{Talk}. To this aim, the approximation of the interaction region in the form of an equivalent magnet with a uniform field with some effective values of the field and length can serve as the simplest model.  In the theory of synchrotron radiation, losses in a such magnet are proportional to the product of the square of its field by the length. When describing the beam-beam interaction a non-zero crossing angle, the doubled size (\ref{sig})
\begin{equation}\label{sig1}
L_{eff}=\frac{4\sigma_x}{\theta}
\end{equation}
can be used, from geometric considerations,  as a full effective length of the 'magnet'.  It is interesting to compare this value with the width of the distribution  of the square of the effective transverse field  $H_\perp$ in Lab, represented by the rate of increase in energy loss during the counter approach of the bunches in the framework of the model 2.  For this purporse,  in the formula for losses (\ref{eLLr}), we perform integration over the variables  $x_1$ and $y$ with the inclusion of the corresponding distribution functions. The remaining integrand, which depends on $\xi$, is normalized to its maximum at $\xi = 0$. As a result, we write the distribution of the loss rate as a function of the distance from the central cross section of the probe bunch to IP: 
\begin{equation}\label{distr}
\frac{H_\perp^2(\xi)}{H_{\perp, max}^2}=\frac{\int\limits_0^\infty\int\limits_0^\infty dtdt'\left[\frac{\sigma_x^2}{\sqrt{g_x^3g_y}}\left(1+\frac{\xi^2\theta^2(2\sigma_x^2+t)(2\sigma_x^2+t')}{\sigma_x^2g_x}\right)+\frac{\sigma_y^2}{\sqrt{g_y^3g_x}}\right]\exp{\left[-\frac{4\xi^2\Omega^2(2\sigma_x^2+t)(2\sigma_x^2+t')}{\sigma_z^2g_x}\right]}}{\int\limits_0^\infty\int\limits_0^\infty dtdt'\left[\frac{\sigma_x^2}{\sqrt{g_x^3g_y}}+\frac{\sigma_y^2}{\sqrt{g_y^3g_x}}\right]}.
\end{equation}
Here, the functionals $g_x(t,t')$, $g_y(t,t')$ and $\Omega(t,t')$ are defined above.  From (\ref{distr})  it can be seen that this characteristic  to a certain extent (rather slightly) depends  on the bunch length $\sigma_z$ (note how $\Omega$ in the exponent depends on $\sigma_z$).

\begin{figure}[htb]
\centering
\includegraphics*[width=100mm]{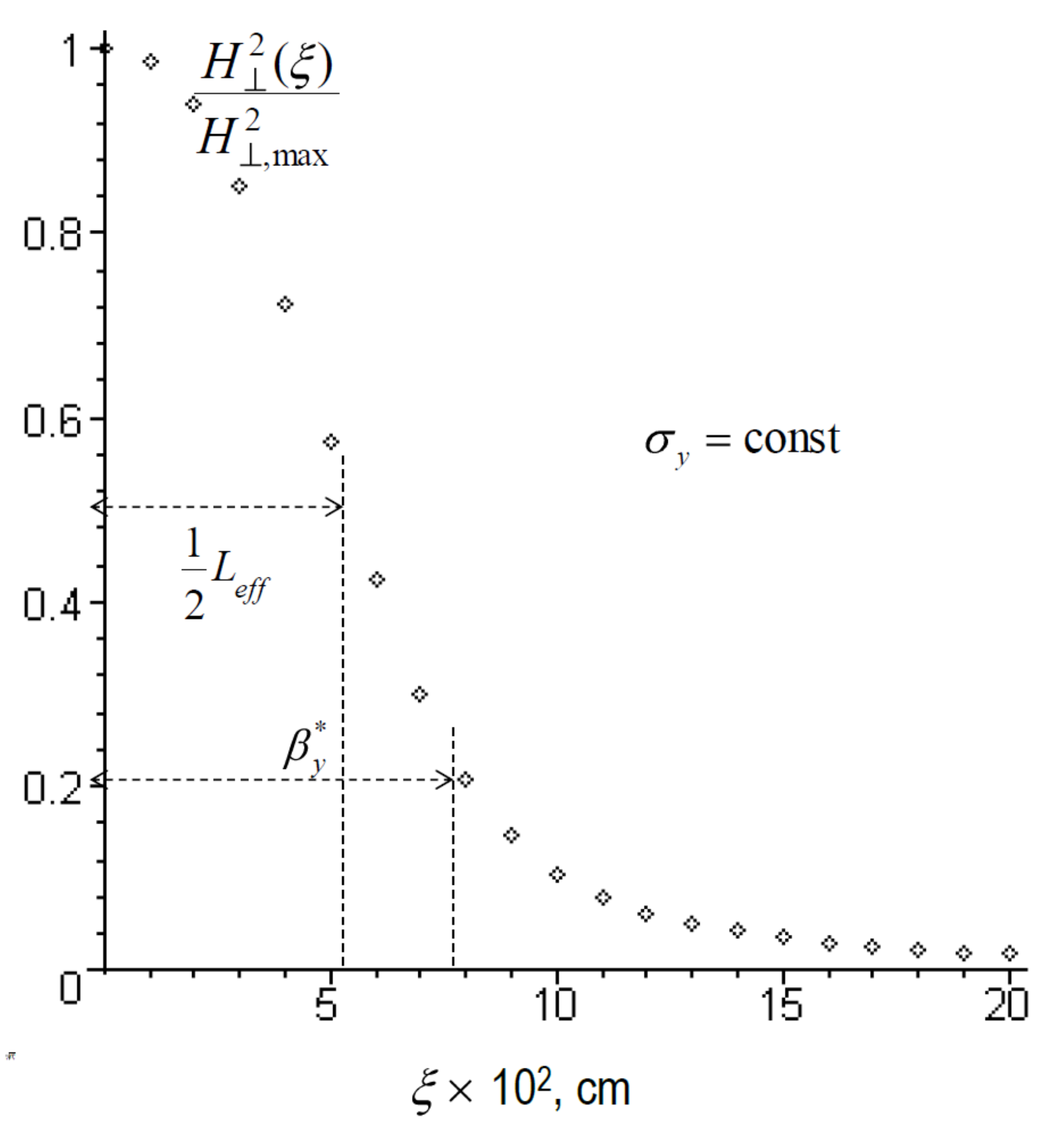}
%\vspace{-15mm}
\caption{The curve of the rate of increase in BS loss with approach of the test bunch to IP in Model 2 as applied to 45 GeV  FCCee case ($\beta_y^*=0.8$ mm).}
\label{bsf4}
\end{figure}

In Fig.\ref{bsf4}, the rate of loss is plotted versus $\xi$  from (\ref{distr}) in a finite range corresponding to moving the centers of bunches up to the IP point. The calculation used the FCCee beam parameters:  $E=45.6$ GeV, $\theta=33$ mrad,  $\sigma_x=6.4$ um, $\sigma_y=28$ nm, $\sigma_z=12.1$ mm \cite{FCC}. Here, the transverse beam sizes are given at IP; the longitudinal size was obtained  by D. Shatilov in the beam-beam simulation taking into account BS. 
The half-height width of the calculated distribution amounts to $\sigma_{1/2}\approx0.055$  cm and corresponds to half the length of the interaction region. It differs little from the same characteristic calculated through the parameter (7) as applied to the Gaussian approximation of the distribution curve: $ \sqrt{-2\ln{0.5}}\frac{2\sigma_x}{\theta}\approx0.051$ cm.

To a certain extent, this result can serve as an indication of the correctness of the obtained formula for energy loss. In addition, it substantiates the choice of quantity (\ref{distr})   as a length of the equivalent magnet.
\begin{figure}[htb]
\centering
\includegraphics*[width=100mm]{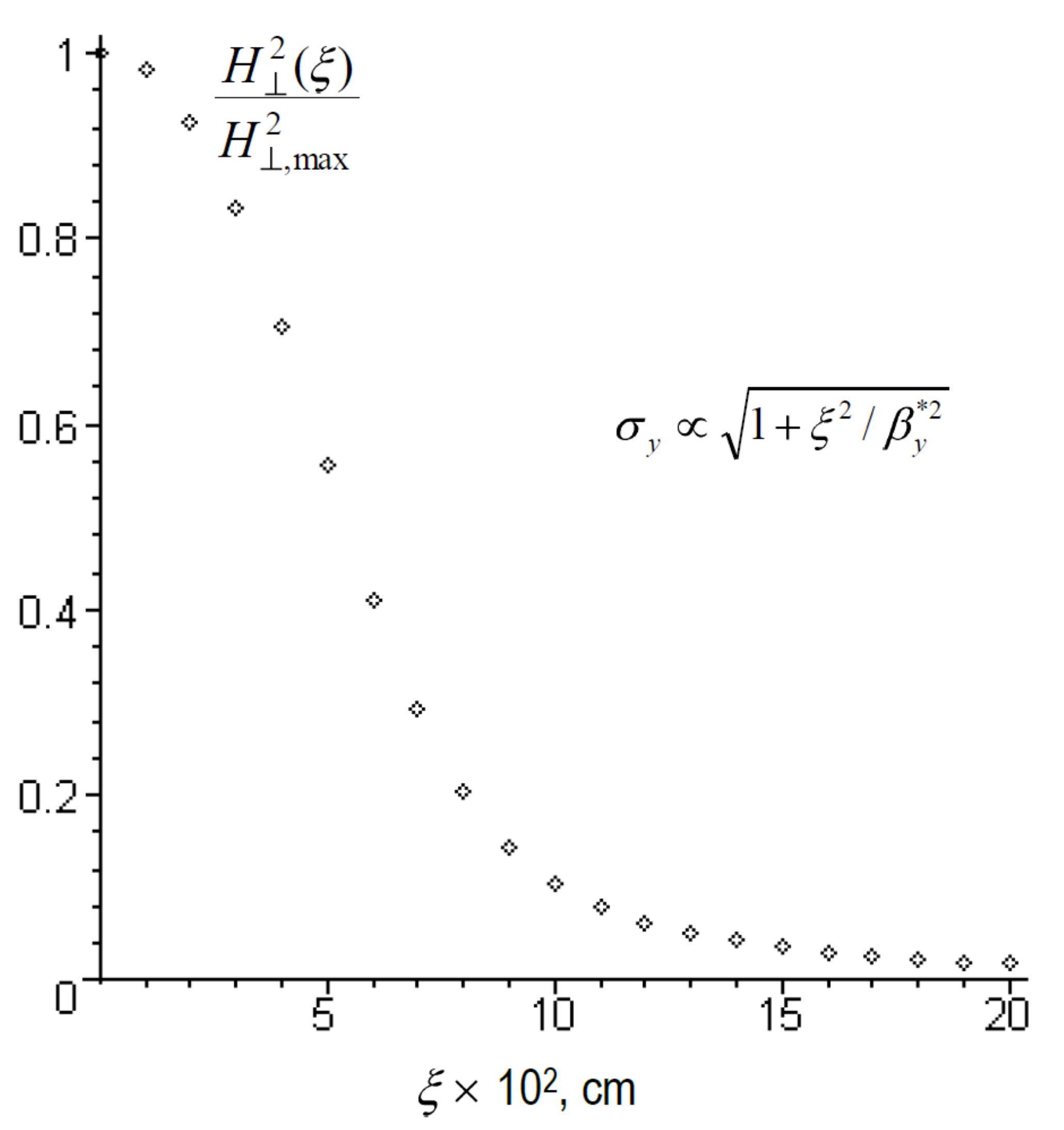}
%\vspace{-15mm}
\caption{The BS loss rate distribution is similar to Fig.\ref{bsf4}, but constructed taking into account the vertical hourglass effect.}
\label{bsf5}
\end{figure}

\section{Discussion}
In our models, we imply that the bunches conserve their shape and sizes.  
In reality, there is an increase of  the transverse bunch sizes at distances comparable 
with or larger than $\beta_y^*$ ($\beta_x^*$), the  vertical (radial) beta function value at IP (hour-glass effect). This occurs 
vertically, starting from rather shorter distances, in comparison with the radial direction, since the vertical beta function is 
much smaller than the radial one.  The function $\beta_y$  grows with increasing $\xi$ as $\beta_y=\beta_y^*+\xi^2/\beta_y^*$, and 
the vertical size $\sigma_y(\xi)\propto \sqrt{\beta_y(\xi)}$.  In the Crab Waist schemes of FCCee anad CEPC  \cite{FCC,CEPC}
\cite{Rai}, $\beta_y^*$ exceeds the characteristic size $0.5 L_{eff}$ of the interaction region (see Fig.\ref{bsf4}). 
At $\xi=0.5 L_{eff}$, the vertical size increases 1.14 times (at 45 GeV FCCee).  Figure \ref{bsf5} shows the dependence of the rate of increase in losses on the distance $\xi$ taking into account the vertical resizing factor due to the corresponding modification (\ref{distr}).
As clearly seen, the difference of the curves in Fig. \ref{bsf4} and Fig. \ref{bsf5} are vanishingly small \footnote{The difference becomes noticeable by artificially increasing the dependence of $\beta_y$ on the distance $\xi$ by several orders of magnitude}. This fact is explained by the peculiarity of the exponent in formula (\ref{distr}).  It tends to its maximum for large values of the integration variables $t, t' \gg \sigma_x^2\gg\sigma_y^2$.   This region makes the main contribution to the integral in the numerator (14), and due to the above inequalities, its dependence on the vertical size disappears.
On this basis, we can conclude that Model 2 is insensitive to the hour-glass effect.
%\begin{figure}[htb]
%\centering
%%\includegraphics*[width=100mm]{pro6958.eps}
%\includegraphics*[width=100mm]{BSfig5.pdf}
%%\vspace{-15mm}
%\caption{The BS loss rate distribution is similar to Fig.\ref{bsf4}, but constructed taking into account the vertical hourglass effect.}
%\label{bsf5}
%\end{figure}

\section{Acknowledgements}
Author thanks Dmitry Shatilov, Mikhail Zobov and Dariya Leshenok for discussions.

\end{document}